
\input phyzzx
\def\PRS{{\it Proc. Roy. Soc. }}
\def\AP{{\it Ann. Phys. }}
\def\half{{1\over 2}}
\def\NP{{\it Nucl. Phys. }}
\def\PL{{\it Phys. Lett. }}
\def\PR{{\it Phys. Rev. }}

\def\PRL{{\it Phys. Rev. Lett.}}

\def\JETP{{\it JETP  }}

\def\JP{{\it J. Phys.}}

\def\e{\epsilon}
\def\nolabels{\def\eqnlabel##1{}\def\eqlabel##1{}\def\reflabel##1{}}
\def\writelabels{\def\eqnlabel##1{%
{\escapechar=` \hfill\rlap{\hskip.09in\string##1}}}%
\def\eqlabel##1{{\escapechar=` \rlap{\hskip.09in\string##1}}}%
\def\reflabel##1{\noexpand\llap{\string\string\string##1\hskip.31in}}}
\nolabels
\global\newcount\meqno \global\meqno=1
\global\meqno=1
\def\eqnn#1{\xdef #1{(\the\meqno)}%
\global\advance\meqno by1\eqnlabel#1}
\def\eqna#1{\xdef #1##1{\hbox{$(\the\meqno##1)$}}%
\global\advance\meqno by1\eqnlabel{#1$\{\}$}}
\def\eqn#1#2{\xdef #1{(\the\meqno)}\global\advance\meqno by1%
$$#2\eqno#1\eqlabel#1$$}
\overfullrule=0pt

\nopagenumbers
\footline={\ifnum\pageno>1\hfil\folio\hfil\else\hfil\fi}
\REF\J{J. Frenkel, {\it Z. Physik} {\bf 37}, 243 (1926).}
\REF\BMT{V. Bargmann, L. Michel and V. L. Telegdi, \PRL
{\bf 2}, 435 (1959).}
\REF\P{A. Papapetrou, \PRS{\bf A209}, 248 (1951).}
\REF\susy{F. Ravndal, \PR{\bf D21}, 2823 (1980).}
\REF\K{I. B. Khriplovich, \JETP{\bf 69}, 217 (1989).}
\REF\holten{J. W. van Holten, {\it\NP}{\bf B356}, 3 (1991).}
\REF\dyson{F. J. Dyson, {\it Am. J. Phys.} {\bf 58}, 209 (1990).}
\REF\lee{C. R. Lee, \PL {\bf A148}, 146 (1990).}
\REF\tan{S. Tanimura, {\it \AP}{\bf 220}, 229 (1992).}
\REF\S{S. K. Soni, \JP {\bf A25}, L837 (1992).}
\REF\SY{A. Stern and I. Yakushin, UAHEP-933 (April, 1993).}
\REF\KB{K. Yee and M. Bander, UCI-TR 93-6 (February, 1993).}
\REF\lan{M. C. Land, N. Shnerb and L. P. Horwitz,
TAUP-2076-93.}
\REF\JN{R.Jackiw and V.P.Nair, {\it Phys.Rev.}, {\bf D43},
1933 (1991).}
\REF\ply{M.S.Plyushchay, {\it Phys.Lett.}, {\bf B248}, 107 (1990);
D.Shon and S.Khlebnikov, {\it JETP Lett.}, {\bf 51}, 611 (1990);
D.Volkov, D.Sorokin and V.Tkach, in {\it Problems in Modern Quantum Field
Theory}, A.Belavin, A.Klimyk and A.Zamolodchikov (eds.) (Springer,
Berlin, 1989).}
\REF\chou{C. Chou, V. P. Nair and A. P. Polychronakos, {\it\PL}
{\bf B304}, 105 (1993).}
\REF\vp{C. Chou and V. P. Nair, (In preparation).}
\REF\c{C. Chou, (In preparation).}
\REF\ch{C. Chou, (unpublished).}

\titlepage
\medskip
\title{Dynamical Equations of Spinning Particles: Feynman's Proof}
\bigskip
\author {Chihong Chou\footnote\dagger
{chou@rock03.rockefeller.edu}}
\address{Physics Department\break
Rockefeller University\break
New York, New York 10021\break
U.S.A.}
\bigskip
\abstract{In this letter, we discuss the extension of Feynman's
derivation of the equation of motion to the case of spinning particles.
We show that a spinning particle interacts only with the electromagnetic
and gravitational fields. In the absence of
the electromagnetic interactions, we rederive
Papapetrou's equations for spinning particles in
the background of the conformal gravity.
We also find that
the effect of spin coupled to non-constant electromagnetic
fields leads to further corrections to
the Lorentz force equations. Some discussions
of these results are given at the end.}
\bigskip
\line{RU93-10-B \hfill}
\line{November 1993 \hfill}
\vfill
\vskip .6in
\noindent
This work was supported in part by the US Department of Energy.
\endpage
\vfill\eject
\baselineskip 24pt

The derivation of the dynamical equations of the spinning particles
in external fields has attracted the interests of physicists
for over fifty years.$^{\J-\holten}$ The goal in attacking
this problem is to study the dynamical effects of the spin precession,
the spin-spin interactions (the Stern-Gerlach effects) and
the spin-orbits couplings for particles in
external fields.  The most notable results of this search
are the Bargmann-Michel-Telegdi (BMT) equations$^\BMT$ and the
Papapetrou equations.$^\P$ Based on what Frenkel suggested,$^\J$
Bargmann, Michel and Telegdi discussed the precession of spinning
particles in external electromagnetic fields,$^\BMT$ while
Papapetrou conjectured the dynamical equations of spinning
particles in general relativity by considering a rotational
mass-energy distribution in the limit of vanishing
volume but with the angular momentum remaining finite.$^\P$
Still various discussions in looking for the
dynamics of spinning particles remain active from
different points of view$^{\K, \holten, \KB, \susy}$
and all of these are less
direct or attractive when compared to Feynman's
rederivation of Lorentz force equations.$^{\dyson}$

Since Dyson$^\dyson$ presented the Feynman's proof
of the homogeneous Maxwell equations and
Lorentz force equation for a Newtonian particle, the generalizations
to the case of the spinless particle
(with and without the internal structure)
in both special and general relativity
have been studied.$^{\lee-\lan}$ The conventional theories
have not been completely rederived. In particular, Tanimura$^\tan$
showed that the particle worldlines are not
parametrized by the proper time of motion
and the particle does not follow a geodesic.
In this paper, we will take a further generalization to the case of
spinning particles and reexamine closely the above mentioned features.
We rederive both the Lorentz force equations and
the Papapetrou equations in a simple and direct way.
Specially, our approach offers a
systematic study of both spin-charge and
spin-gravitational coupling to all orders.

By postulating the Poisson brackets
of the Newtonian variables of particle motion, Feynman, according to
Dyson, obtained the homogeneous Maxwell equations and the Lorentz
force equations.$^\dyson$ The key points in his proof are
to use the associative conditions, namely the Jacobi identity
of the brackets, and the so-called second Leibniz rule:$^\tan$
\eqn\L{{d\over d\tau}[A, B]=
[{dA\over d\tau}, B]+[A, {dB\over d\tau}],}
where the $\tau$ is a parameter of the particle trajectory.
Note that using the Jacobi identity
one may derive Eq. \L \ if one assumes the
existence of Hamiltonian evolution.
Therefore, this procedure
may be re-formulated in terms of the symplectic language, in which
the symplectic two-form, by definition yields the
particle's Poisson brackets and their associated Jacobi identity.
Moreover, the homogeneous Maxwell equations follow trivially from
the closure of the symplectic two-form
and the Lorentz force equations may be derived
simply from one or two line computations. We shall use this
new formulation of the Feynman's approach to study spinning particles.

Since the spin degree of freedom arises from symmetry transformations
of space-time, it is generally not possible to formulate the spin variables
purely in terms of Newtonian coordinates of
the particle motion. This difficulty prevents one from directly
employing Feynman's proof.
However, in two spatial dimensions, the canonical structure of
a spinning particle is explicitly known$^{\JN, \ply, \chou}$
and the spin degree of freedom
indeed can be purely  represented by the Newtonian variables of the
particle motion.$^{\JN,\chou}$

We shall begin with a discussion of a
spinning charged particle in 2+1 dimensional
flat space-time. In this paper, we assume the particle
without internal structure (extension to the case of internal
degrees of freedom is straightforward).
For a spinning particle or an anyon with spin $-s$ in a
2+1 dimensional flat space-time, with the metric $\eta_{ab}=diag(+--)$,
the symplectic structure is given by$^{\chou}$
\eqn\f{\omega=dx^a\wedge dp_a+ {{1\over 2}}s f_{ab}dp^a\wedge dp^b
+ {{1\over 2}}eF_{ab}dx^a\wedge dx^b.}
Where $p^a=m{\dot{ x}}^a$, $p^2=\eta_{ab}p^ap^b$ ($\e_{012}=1$),
$f_{ab}=\e_{abc}p^c/(p^2)^{3/2}$, and $x^a(a=0,1,2)$
are the position variables of a particle(the overdot denotes the
$\tau$-derivative and $m$ is the particle's mass). The spin
vector $S^a$ for the particles, as shown in Ref. [\JN, \chou], is given by
\eqn\S{S^a=-s{p^a\over \sqrt{p^2}}.}

The closure condition $d\omega=0$ tells us that the antisymmetric tensor
$F_{ab}$ is a function of $x^a$ only and satisfies the homogeneous Maxwell
equations:
\eqn\ma{\partial_c F_{ab} +
\partial_a F_{bc} + \partial_b F_{ca} =0.}
By definition, $\omega$ gives the Poisson brackets or the commutation
rules as
\eqn\q{\eqalign{&[x^a, \ x^b]=is({\widetilde M}^{-1}f)^{ab}, \cr
	        &[p^a, \ x^b]=i(M^{-1})^{ab}, \cr
		&[p^a, \ p^b]=ie(M^{-1}F)^{ab},}}
where $M_{ab}=\eta_{ab}+ es(Ff)_{ab}~\equiv
\eta_{ab}+es F_{ac}f^c_{~b}$. $\widetilde M$ denotes the
transpose of $M$.

Using the second Leibniz rule \L \ and Eqs. \q,
we have,
\eqn\WW{\eqalign{{d\over d\tau}({M}^{-1})^{ab}
&=i^{-1}\left({1\over m}[p^a,\ p^b]+[{\dot{ p}}^a,\ x^b]\right), \cr
&={1\over m}(M^{-1}F)^{ab}+i^{-1}[{\dot {p}}^a,\ x^b].}}
In principle, one may deduce from Eq. \WW \
the equation of motion of
the spinning particle for arbitrary
value of spins and external field $F$.
It is, of course, very difficult to simplify these equations
in general to obtain the desired equations.
We shall instead derive the equations of motion
in terms of a series in powers of spin.
Since the $\tau$ derivative of $p^a$ is also hidden in the left
hand side of Eq. \WW, it is easy to deduce the
equations of motion from the first equation of \q \ by
applying the Leibniz rule \L. Thus
\eqn\W{\eqalign{ms{d\over d\tau}({\widetilde M}^{-1}f)^{ab}
&=i^{-1}\left([p^a,\ x^b]+[x^a,\ p^b]\right), \cr
&=(M^{-1})^{ab}-(M^{-1})^{ba}.}}

We expand Eq. \W \ into a power series of $s$ and keep
only the lowest nontrivial order:
\eqn\WW{\eqalign{{d f^{ab} \over d\tau}
=&{e\over m}\left[(Ff)^{ba}-
(Ff)^{ab}\right]+es{d\over d\tau}(fFf)^{ab} \cr
&+{e^2s\over m}\left[(FfFf)^{ab}-(FfFf)^{ba}\right]
+{\cal O}(s^2), }}
which may be further simplified into
\eqn\j{{dY^a\over d\tau}=-{e\over m}F^{ab}Y_b
-{es\over m}Y^a Y\cdot \dot{F} +~{\cal O}(s^2),}
where $Y^a={1\over 2}\e^{abc}f_{bc}$ and $F_a={1\over 2}\e_{abc}F^{bc}$.
Thus, we have the equations of motion for both the position
and spin variables (3):
\eqn\EEE{\eqalign{{dp^a\over d\tau}+{e\over m}F^{ab}p_b&=
{3\over 2} p^a{d\over d\tau}\ln(p^2)+
{e\over m}{p^a\over p^2}S^b({\partial^c F_b})p_c
+ {\cal O}(s^2) \cr
&=-{e\over 2 m}{p^a\over p^2}S^b({\partial^c F_b})p_c
+ {\cal O}(s^2), }}
\eqn\S{{dS^a\over d\tau}=-{e\over m}F^{ab}S_b+{\cal O}(s^2).}

Clearly, Eq. \S \ is the Bargmann-Michel-Telegdi
equations in 2+1 dimensions.$^{\BMT, \chou}$
Since $\partial F$ is always coupled to the spin,
$\partial F$ term will not give any contributions to
order $s^2$ to the equations
of motion for the spin. We will see this again
below. However, the equations \EEE \ are physically wrong although
it follows from our starting point. First of all, they are
not the Hamiltonian equations, because the $\tau$-evolution
is not generated by the $\tau$-Hamiltonian, which is proportional to
the first integral of motion of Eqs. \EEE, i.e. $A=p^2+eS\cdot F$.
Secondly, Eqs. \EEE \ are not the typical
Hamiltonian equations of motion
because of explicit $\tau$-derivative appearing
in the right hand side of the equations.
Alternatively speaking, using the rules
in Eq. \q \ we cannot derive
Eqs. \EEE \ and their counterpart $m\dot{x}^a=p^a$
as the Heisenberg equations.
For example, $\dot{x}^a\ne i^{-1}[x^a, \zeta A]$ for any
constant $\zeta$. This apparent inconsistency
arises as the consequence of the spin
coupled to the gradient of the external electromagnetic
field, although it is perfectly consistent
for a slowly varying field $F$, as we will demonstrated below.

For a slowly varying field $F$, Eqs. \S \ remain unchanged and
Eqs. \EEE \ yield:
\eqn\E{\eqalign{&{d\over d\tau} p^2 +~{\cal O}(\partial F, s^2)=0, \cr
	&{dp^a\over d\tau}=-{e\over m}F^{ab}p_b
		+~{\cal O}(\partial F, s^2),}}
i.e. we have the (correct) worldline-equation
and the Lorentz force equations
(upto possible terms which depend on gradients of $F$).
As shown in Ref. [\chou], the $\tau$-Hamiltonian $G$
which generates $\tau$-evolution for this system can be obtained
from \f \ and \E:
\eqn\H{G=-{1\over 2m}\left(p^2-2e S\cdot F-m^2\right)
+{\cal O}(s^2).}
Notice that Eq. \H \ is not the first integral of motion of \E,
but rather is that of a similar equation:$^{\holten, \vp}$
\eqn\EE{{dp^a\over d\tau}+{e\over m}F^{ab}p_b=
{e\over m}S^b{\partial^a F_b} + {\cal O}(s^2).}
These equations in Eqs. \E \ or \EE \ are indeed realized quantum
mechanically as the Heisenberg equations:$^{\chou}$
\eqn\HHe{{d\eta^a \over d\tau}={1\over i}[\eta^a, G],}
where $\eta^a=(x^a, p^a)$.

Notice that one needs only the rules
in Eq. \q \ to order $s$ to verify the equivalence between
Eq. \HHe \ and \EE. However, when we used the first equation in
Eq. \q \ to derive Eq. \EEE, we actually use the rules
in Eq. \q \ to order $s^2$. This suggests that the postulated
symplectic structure or the Poisson brackets in Eq. \q \ are
not compatible with the definition of $m \dot{x}^a\equiv p^a$
to order $s^2$. In other word, it is necessary to either find the
proper higher order corrections to $\omega$ or modify the relation
$m \dot{x}^a\equiv p^a$ to resolve the physically unacceptable
situation mentioned above. In fact, one may find that if
one assumes
$m \dot{x}^a=p^a+ems k^a\equiv p^a+ems\e^{abc}\partial_b(S\cdot F)Y_c$,
in stead of having the equation (7) or (10),
one has a new equation by applying the rule \L \
\eqn\WW{\eqalign{ms{d\over d\tau}({\widetilde M}^{-1}f)^{ab}
&=i^{-1}\left([p^a,\ x^b]+[x^a,\ p^b]+
{ems}\{[k^a, x^b]+[x^a, k^b]\}\right), \cr
&=(M^{-1})^{ab}-(M^{-1})^{ba}
+{ems}\left({\partial k^a\over\partial p_b}-
{\partial k^b\over\partial p_a}\right)
+{\cal O}(s^3).}}
Similarly, we have
\eqn\jj{{dY^a\over d\tau}=-{e\over m}F^{ab}Y_b
+{es\over m}\left[3Y^a Y\cdot \dot{F}-
p\cdot Y \partial^a(Y\cdot F)\right] +~{\cal O}(s^2).}
One may easily show that Eq. \jj \ leads to
both the Hamiltonian in Eq. \H \ and Eq. \EE \ or \E.

Equivalently, one can keep the relation $m{\dot{x}}^a=p^a$ and
change the symplectic structure
to
\eqn\ff{{\widetilde{\omega}}=dx^a\wedge dp_a+
{{1\over 2}}s f_{ab}dp^a\wedge dp^b
+ {{1\over 2}}eF_{ab}dx^a\wedge dx^b
- emsd x^a \wedge d k_a,}
which is compatible with the relation $m \dot{x}^a=p^a$
to order $s^2$. In general, it is very difficult to find
the (exact) symplectic structure which is compatible with
$m \dot{x}^a\equiv p^a$ to arbitrary order of s. However,
what we have provided is a systematic way of constructing
such model. Namely, we can repeatedly do the above procedure.

Now we consider the case of the space-time with metric $g_{ab}(x)$.
To avoid lengthy computations, we first consider the case
of pure gravitational interactions. The (simplest) symplectic structure
will take the following form
(the numeral tensor density $\e_{abc}$ is assumed):$^{\vp}$
\eqn\i{\omega_g=dx^a\wedge dp_a + {{1\over 2}}s\sqrt{g}
f_{ab} dp^a\wedge dp^b
+{1\over 2}dx^a\wedge d(S_{a}^b\partial_b\ln p^2),}
where $g=det g_{ab}$, $S_{ab}=-s\sqrt{g} p^2f_{ab}$, and the index
is raised or lowered by $g_{ab}$. The closure
condition of $\omega_g$ yields
\eqn\G{(p^2g^{ab}-3p^ap^b)dg_{ab}=0,}
which is satisfied only by
the conformal flat metric, where $g^{ab}$ is the inverse of $g_{ab}$.

For a conformal flat metric $g_{ab}=\eta_{ab}e^{\phi(x)}$,
we may rewrite $\omega$
in term of $x^a$, $u^a=m\dot{x}^a$:
\eqn\d{\eqalign{\omega_g&=e^\phi\left({\tilde{\eta}_{ab}}dx^a\wedge du^b
-{{1\over 2}}u^2S_{ab}(u)du^a\wedge du^b
+ {1\over 2}E_{ab}dx^a\wedge dx^b\right),}}
where
\eqn\dd{\eqalign{&\widetilde{\eta}_{ab}=\eta_{ab}+{1\over 2}{\partial S_{ac}
\over \partial u^b}\phi^c \cr
&E_{ab}=u_a\phi_b-u_b\phi_a+{1\over 2}(S_{ac}\phi^c_b-S_{bc}\phi^c_a) \cr
&S^{ab}=\e^{abc}S_c\exp(-\phi), \ \ \ \ \phi_a=\partial_a\phi,
\ \ \ \ S_a=-s{u_a\over \sqrt{u^2}}.}}
Notice $S^{ab}$ is the tensor with respect to the metric $g_{ab}$
and the index is lowered or raised
by the flat metric $\eta_{ab}$ and we will use this
convention from now on except for the explicit indication.

The two-form in Eq. \d \ gives the commutation rules:
\eqn\qq{\eqalign{&[x^a, \ x^b]=ise^{-2\phi}({\tilde\eta}^{-1})^{ca}f_{cd}
({\tilde N}^{-1})^{bd}, \cr
&[u^a, \ x^b]=ie^{-\phi}(N^{-1})^{ab}, \cr
&[u^a, \ u^b]=ie^{-2\phi} ({\tilde\eta}^{-1})^{ac}E_{cd}(N^{-1})^{bd},}}
where $N_{ab}={\tilde \eta}_{ab}+ sE_{ac}({\tilde\eta}^{-1})^{dc}f_{db}
\exp(-\phi)$ and
$\tilde N$ denotes the transpose of $N$.
Again, we seek an expansion in term of a series of spin. The first two of
these rules (all one needs) become:
\eqn\qqq{\eqalign{&[x^a, \ x^b]=ise^{-2\phi}f_{ab}
+\ {\cal O}(s^3) \cr
&[u^a, \ x^b]=ie^{-\phi}(\eta^{ab}+K^{ab} +D^{ab})
+\ {\cal O}(s^3),}}
where
\eqn\qr{\eqalign{
&K^{ab}=-{1\over 2}{\partial S^{ac}\over
\partial u_b}\phi_c+{1\over u^2}u^a\phi_c S^{cb} \cr
&D^{ab}=(KK)^{ab}+{1\over 2 u^2}\left(
(SS)^{bc}(\phi^a\phi_c+\phi_c^a)+S^{ad}\phi_{dc}S^{cb}\right).}}

Applying the rule \L \ on the first equation in Eq. \qqq \ and using
the second one in the same equations, we have
\eqn\pe{\dot{Y}^a-2Y^a\dot{\phi}={1\over s}\e^{abc}
(K_{bc}+D_{bc})+{\cal O}(s^2),}
which may be simplified into
\eqn\pee{\dot{Y}^a-\half (u\cdot Y\phi^a+
Y^a\dot{\phi})={1\over 2(u^2)^{3\over 2}}S^{ab}
\left(\half \phi_b \dot{\phi}-{\dot{\phi}}_b\right)
+{\cal O}(s^2).}
Immediately, we may deduce from the equations \pee
\eqn\H{\eqalign{&{d\over d\tau}\left(u^2 e^{\phi}\right)
+{\cal O}(s^2)=0 \cr
&{du^a\over d\tau}-\half (u^2\phi^a+
u^a\dot{\phi})={1\over 2}S^{ab}
\left(\half \phi_b \dot{\phi}-{\dot{\phi}}_b\right)
+{\cal O}(s^2).}}

The $\tau$-Hamiltonian $G_g$, which is proportional to
the first integral of motion, may be determined by the method
shown in Ref. [\chou] and a similar calculation yields:
\eqn\g{G_g=-{1\over 2m}\left(u^2 e^\phi-m^2\right) + {\cal O}(s^2).}
One may check that Eqs. \H \ are indeed the $\tau$-Hamiltonian
equations of motion by using the commutation rules \qq \ and
the Heisenberg equations:
\eqn\He{\eqalign{&{d x^a \over d\tau}={1\over i}[x^a, G_g] \cr
&{du^a \over d\tau}={1\over i}[u^a, G_g].}}

Recalling the forms of the Christoffel symbol and Riemann tensor
in the case of the conformal metric:
\eqn\cq{\eqalign{&\Gamma^a_{bc}=\half(\delta_c^a{\phi}_b
+\delta_b^a{\phi}_c-\eta_{bc}\phi^a) \cr
&R^a_{bcd}={1\over 4}\delta_d^a({\phi}_b\phi_c-\eta_{bc}\phi^m\phi_m
-2\phi_{bc}) + {1\over 4}\eta_{bc}(\phi^a\phi_d-2 \phi^a_d) \cr
& \ \ \ \ \ \ \ \ -(d\leftrightarrow c),}}
we can rewrite the equations of motion in Eqs. \H \
in a more familar form,
\eqn\pee{{d u^a\over d\tau}+ {1\over m} \Gamma^a_{bc}u^b u^c+\half
R^a_{bcd}u^bS^{cd}={\cal O}(s^2).}
The equations of motion for the spin may also be obtained,
\eqn\pes{{D S^{ab}\over D\tau}\equiv
{d S^{ab}\over d\tau}+{1\over m} \Gamma^a_{cd}S^{cb}u^d
+{1\over m} \Gamma^b_{cd}S^{ac}u^d ={\cal O}(s^2).}
One recognizes that Eqs. \pee \ and \pes \ are in fact the
Papapetrou equations at 2+1 dimensions.$^{\P, \vp}$
In particular, Eqs. \pee \ is
formally similar to the Lorentz force law \E, in which
the field strength $F_{ab}$, the scalar charge $e$
are replaced by the space-time curvature, the
tensorial coupling $S^{ab}$, respectively, while
$S^{ab}$ is covariantly constant of motion.

For the case of including the external electromagnetic
field, the computation is straightforward but lengthy,
and we here only state the results.
The symplectic structure for this case is
\eqn\EM{\omega_{new}=\omega_g+\half F_{ab} dx^a\wedge dx^b
-ems dx^a\wedge d k_a,}
where $k_a$ is given by
\eqn\k{k_a=\half \e_{abc}S^{mn}{DF_{mn} \over Dx_b}Y^c
e^{-\phi},}
and
\eqn\D{\eqalign{&{D F_{mn} \over D x^l}=
{\partial F_{mn} \over \partial x^l}-\Gamma^k_{lm}F_{kn}
-\Gamma^k_{ln}F_{mk} \cr
&\half \e^{mnb}{DF_{mn}\over Dx_a}
=\partial^a F^b-F^b\phi^a-\half F^a \phi^b+
\half F^c\phi_c \eta^{ab}.}}

The equations of motion for both the particle's positions
and spin variables are:
\eqn\pee{\eqalign{&{d u^a\over d\tau}+{1\over m}
\Gamma^a_{bc}u^b u^c+{e\over m}F^{ab}u_b e^{-\phi}+\half
R^a_{bcd}u^bS^{cd}\cr
&\ \ \ \ ={e\over 2m} e^{-\phi}
S^{cd}{D F_{cd} \over D x_a}
+ {\cal O}(s^2) \cr
&{d S^{ab}\over d\tau}+{1\over m} \Gamma^a_{cd}S^{cb}u^d
+{1\over m} \Gamma^b_{cd}S^{ac}u^d
=-{e\over m}\e^{abc}F_{cd}S^d e^{-2\phi} + {\cal O}(s^2).}}
Note that these equations in Eq. \pee \ are obviously
generally covariant and the first one has been obtained
previously.$^{\holten, \vp}$
Again, the equations \pee \ are
consistent with the (appropriate) Heisenberg equations
(Hamiltonian evolution) as mentioned above,
where the $\tau$-Hamiltonian is
\eqn\HH{G_{new}=-{1\over 2m}\left(u^2 e^\phi-
2e S\cdot F e^{-\phi}-m^2\right) +{\cal O}(s^2).}

Some remarks are in order. If one postulates the existence of
Hamiltonian equations of motion, one may easily obtain
the desired equations by finding the Hamiltonian $G$ from the relation
$m\dot{x}^a=i^{-1}[x^a, G]=p^a$.$^{\vp}$ Namely, if one proves that the second
Leibniz rule is equivalent to the existence of Hamiltonian
equations of motion, Feynman's proof will not be
necessary. But with use of second Leibniz rule(Feynman's ideas),
it is practically easier to determine the dynamics of spinning particles
if one is interesting in the higher order spin-spin interactions.
We are able to carry out a general proof that the second
Leibniz rule does indeed lead to the existence of
the Hamiltonian evolution. This proof and its application
for constructing the dynamics of spinning particle
in $3+1$ dimensional space-time
will be published elsewhere.$^{\c}$

As demonstrated above, the inconsistency arises only in the case
of that the spin is coupled to the fast varying
electromagnetic field, i.e. the gradients of the external field
strength $F$. This is due to the fact that the symplectic structure
for spinning particles is not exactly known and what we had
is not compatible with the postulated relation $m \dot{x}^a
\equiv p^a$. We believe that with proper modification
of the symplectic structure of spinning particles, we can
completely resolve this matter. We have shown that starting with
modifying the relation between the phase space variables $p^a$
and the velocities of particles, the second Leibniz rule
leads to desired equations, which are also Heisenberg equations.
This procedure demonstrates
a systematic method to study the dynamical effects of
spin coupled to the external fields to any order $s$.

In the absence of the electromagnetic interactions, the geodesic
equation for particle's position (with the standard spin-curvature
term of Papapetrou$^{3}$) is obviously valid. This feature
was not rediscovered in Ref. [\tan] as the consequence
of allowing for interactions with external scalar field. This is because
if a relativistic particle interacts with
the scalar field, its position will not follow a geodesic. The
argument offered by Tanimura$^{\tan}$ is irrelevant, since
the geodesic condition $g_{ab} \dot{x}^a \dot{x}^b=1$ appears
as the constraint equation and it will require Dirac's
procedure to further test its consistency with
the postulated Poisson brackets. Our parameter
$\tau$ may be interpreted as the proper time of particle's motion.

We have demonstrated that the only possible fields that can consistently
act on a quantum mechanical spinning particle in 2+1 dimensions
are gauge and gravitational fields. Therefore,
there will be a lack of continuity
when we take the limit of spin going to zero, as we have learned
from Ref. [\tan], where the scalar interactions may be allowed.
In other words, a spinless particle may not be viewed as
a spinning particle at the limit of $s\rightarrow 0$, because
of possible scalar interactions.
Finally, since our derivation is limited to the case of
the conformal gravitational backgrounds, it will be
interesting to see if one can carry out the same analysis for
a general gravitational field; this will also shed light on
how an anyon interacts with the gravitational fields.
Generalization to 3+1 dimensions will be discussed
elsewhere.$^{\c}$ Although the spin is no longer relevant
to the 1+1 dimensional space-time, a special parameter
arises in a similar position to the spin at 2+1. When we
apply our approach to this case, we find even more
peculiar features.$^{\ch}$

\vskip .2in
\ack{We thank V. P. Nair for many valuable suggestions
and discussions, Mark Evans, N. N. Khuri, and H. C. Ren
for discussions, and R. Jackiw and V. P. Nair for
a critical reading of the manuscript.}
\refout
\end